\documentstyle[aps,preprint,amssymb,floats,graphicx]{revtex}

\newcommand{\e}{{\rm e}}                  
\newcommand{\Pomeron }{\mathbb{P}}        
\newcommand{\T}[1]{{\bf #1}_{T}}          
\newcommand{\betaH}{\beta_{\rm H}}        

\begin{document}
\draft
\tighten

\preprint{
        \parbox{1.5in}{%
           \noindent
           hep-ph/9705393 
        }
}

\title{Light-cone Variables, Rapidity and All That}

\author{John C. Collins}
\address{Penn State University,
        104 Davey Lab, University Park PA 16802, U.S.A.
}

\date{22 May 1997}

\maketitle
\begin{abstract}
    I give a pedagogical summary of the methods of light-cone variables,
    rapidity and pseudo-rapidity.  Then I show how these methods
    are useful in analyzing diffractive high-energy collisions.
\end{abstract}



\section{Introduction}
\label{sec:introduction}

The use of light-cone variables, rapidity and pseudo-rapidity is
very common in treating high-energy scattering, particularly
in hadron-hadron and lepton-hadron collisions.  The essential
features of these collisions that make these variables of utility
are the presence of ultra-relativistic particles and a preferred
axis.

In these notes I will explain the theory of these variables.  The
only prerequisites are a knowledge of the elements of special
relativity and an acquaintance with the general phenomenological
features of high-energy scattering.  Then I will show how these
variables can be used conveniently in the analysis of the
kinematics of diffractive processes.  This will include the
derivation of an approximate relation between the size of the
rapidity gap and the variable $x_{\Pomeron }$.

The material in these notes is essentially all standard and
well known to many workers in the field.  But it is not always
easy to find the material conveniently presented.

\section{Light-cone coordinates, rapidity}

\subsection{Definitions}

Light-cone coordinates are defined by a change of variables from
the usual $(t,x,y,z)$ (or $(0,1,2,3)$) coordinates.  Given a
vector $V^{\mu }$, its light-cone components are defined by
\begin{eqnarray}
    V^{+} = \frac {V^{0}+V^{3}}{\sqrt 2} ,
{}~~~
    V^{-} = \frac {V^{0}-V^{3}}{\sqrt 2} ,
{}~~~
    {\bf V}^{T} = (V^{1},V^{2}) ,
\label{def}
\end{eqnarray}
and I will write $V^{\mu }=(V^{+},V^{-},{\bf V}^{T})$.  Some authors prefer to
omit the $1/\sqrt 2$ factor in Eq.\ (\ref{def}).  It can easily be
verified that Lorentz invariant scalar products have the
form\footnote{
    I prefer to make a distinction between contravariant vectors,
    whose indices are superscripts, and covariant vectors, whose
    indices are subscripts.  In all contexts where the
    distinction matters I will put the Lorentz indices in their
    correct places.
}
\begin{eqnarray}
    V\cdot W &=& V^{+}W^{-} + V^{-}W^{+} - \T V\cdot \T W ,
\nonumber\\
    V\cdot V &=& 2V^{+}V^{-} - V_{T}^{2} .
\label{scalar.prod}
\end{eqnarray}

What are the motivations for defining such coordinates, which
evidently depend on a particular choice of the $z$ axis?
One is that these coordinates transform very simply under boosts
along the $z$-axis.
Another is that when a vector is highly boosted along the $z$
axis, light-cone coordinates nicely show what are the large and
small components of momentum.
Typically one uses light-cone coordinates in a situation like
high-energy hadron scattering.
In that situation, there is a natural choice of an axis, the
collision axis, and one frequently needs to transform between
different frames related by boosts along the axis.
Commonly used frames include the rest frame of one of the
incoming particles, the overall center-of-mass frame, and the
center-of-mass of a partonic subprocess.

\subsection{Boosts}

Let us boost the coordinates in the $z$ direction to make a new
vector $V'^{\mu }$.  In the ordinary $(t,x,y,z)$ components we have the
well known formulae
\begin{eqnarray}
    V'^{0} = \frac {V^{0} + v V^{z}}{\sqrt {1-v^{2}}} ,
{}~~~
    V'^{z} = \frac {v V^{0} + V^{z}}{\sqrt {1-v^{2}}} ,
{}~~~
    V'^{x} = V^{x} ,
{}~~~
    V'^{y} = V^{y} .
\end{eqnarray}
It is easy to derive the following for the light-cone components:
\begin{eqnarray}
    V'^{+} = V^{+} \e^{\psi } ,
{}~~~
    V'^{-} = V^{-} \e^{-\psi } ,
{}~~~
    \T{V'} = \T V ,
\end{eqnarray}
where the hyperbolic angle $\psi $ is $\frac {1}{2} \ln \frac {1+v}{1-v}$, so
that
$v = \tanh \psi $.

Notice that if we apply two boosts of parameters $\psi _{1}$ and $\psi _{2}$
the result is a boost $\psi _{1}+\psi _{2}$.  This is clearly simpler than the
corresponding result expressed in terms of velocities.

\subsection{Rapidity}

\paragraph{Boost of particle momentum}
Consider a particle of mass $m$ that is obtained by a boost $\psi $
from the rest frame.  Its momentum is
\begin{eqnarray}
    p^{\mu } &=& \left( p^{+}, \frac {m^{2}}{2p^{+}}, \T 0 \right)
\nonumber\\
       &=& \left( \frac {m}{\sqrt 2} \e^{\psi }, \frac {m}{\sqrt 2} \e^{-\psi
}, \T 0
           \right)
\end{eqnarray}
Notice that if the boost is very large (positive or negative),
only one of the two non-zero light-cone components of $p^{\mu }$ is
large; the other component becomes small. With the usual
components two of the components, $p^{0}$ and $p^{z}$, become large.

Suppose next that we have two such particles, $p_{1}$ and $p_{2}$, with
the boost for particle 1 being much larger than that for particle
2.  Then in the scalar product of the two momenta only one
component of each momentum dominates the result, thus for example
$(p_{1}+p_{2})^{2} \simeq 2p_{1}^{+}p_{2}^{-}$.  This implies that,
when analyzing the sizes of scalar products of highly boosted
particles, it is simpler to use light-cone components than to use
conventional components.

\paragraph{Definition of rapidity}
Since the ratio $p^{+}/p^{-}$ gives a measure $\e^{2\psi }$ of the boost from
the rest frame, we are led to the following definition of a
quantity called ``rapidity'':
\begin{equation}
   y = \frac {1}{2} \ln \frac {p^{+}}{p^{-}} = \frac {1}{2} \ln \frac {
E+p^{z}}{E-p^{z}} ,
\end{equation}
which can be applied to a particle of non-zero transverse
momentum.  The 4-momentum of a particle of rapidity $y$ and
transverse momentum $\T p$ is
\begin{equation}
   p^{\mu } = \left(
           \sqrt {\frac {m^{2}+p_{T}^{2}}{2}} \, \e^{y},
           \sqrt {\frac {m^{2}+p_{T}^{2}}{2}} \, \e^{-y},
           \T p
        \right) ,
\label{particle}
\end{equation}
with $\sqrt {m^{2}+p_{T}^{2}}$ being called the transverse energy $E_{T}$ of
the
particle.  It can be checked that the scalar product of two
momenta is
\begin{equation}
   p_{1}\cdot p_{2} = E_{1T} E_{2T} \cosh (y_{1}-y_{2}) - {\bf p}_{1T}\cdot
{\bf p}_{2T} .
\end{equation}
In the case where the transverse momenta are negligible, this
reduces to $m^{2}\cosh(y_{1}-y_{2})$, which is like the formula for the
product of two Euclidean vectors
${\bf p}_{1}\cdot {\bf p}_{2} = p_{1}p_{2}\cos\theta $, with the trigonometric
cosine
being replaced by the hyperbolic cosine.

\paragraph{Transformation under boosts}
Under a boost in the $z$ direction, rapidity transforms
additively:
\begin{equation}
   y \to  y' = y + \psi  .
\label{boost.of.y}
\end{equation}
This implies that in situations where we have a frequent need to
work with boosts along the $z$ axis it is economical to label
the momentum of a particle by its rapidity and transverse
momentum, rather than to use 3-momentum.

\paragraph{Rapidity distributions in high-energy collisions}
It also happens that in most collisions in high-energy hadronic
scattering, the distribution of final-state hadrons is
approximately uniform in rapidity, within kinematic limits.  That
is, the distribution of final-state hadrons is approximately
invariant under boosts in the $z$ direction.  This implies that
rapidity and transverse momentum are appropriate variables for
analyzing data and that detector elements should be approximately
uniformly spaced in rapidity.  (What is physically possible is to
make a detector uniform in the angular variable pseudo-rapidity
that I will discuss below.)  This is in contrast to the situation
for $e^{+}e^{-}$ collisions where most of the interest is in events
generated via annihilation into an electro-weak boson. Such
events are much closer to uniform in solid angle than uniform in
rapidity.

\paragraph{Non-relativistic limit}
Observe that for a non-relativistic particle, rapidity is the
same as velocity along the $z$-axis, for then
\begin{eqnarray}
    y &=& \frac {1}{2} \ln \frac {E+p^{z}}{E-p^{z}}
       \simeq \frac {1}{2} \ln \frac {m+mv^{z}}{m-mv^{z}}
       = v^{z} .
\end{eqnarray}
Non-relativistic velocities transform additively under boosts,
and the non-linear change of variable from velocity to rapidity
allows this additive rule Eq.\ (\ref{boost.of.y}) to apply to
relativistic particles (but only in one direction of boost).

One way of seeing this is as follows: The relativistic law for
addition of velocities in one dimension is
\begin{equation}
   \beta _{13} = \frac {\beta _{12}+\beta _{23}}{1 + \beta _{12}\beta _{23}},
\label{rel.add}
\end{equation}
where $\beta _{12}$ is the velocity of some object 1 measured in the
rest-frame of object 2, etc.  This formula is reminiscent of the
following property of hyperbolic tangents:
\begin{equation}
   \tanh (A+B) = \frac {\tanh A + \tanh B}{1 + \tanh A \tanh B} .
\end{equation}
So to obtain a linear addition law, we should write $\beta _{12} = \tanh
A_{12}$, and then the rule Eq.\ (\ref{rel.add}) for the addition of
velocities becomes simply $A_{13}=A_{12}+A_{23}$.  The $A$ variables
are exactly relative rapidities, since 
\begin{equation}
  v^z = \frac{p^z}{E} = \frac{p^+-p^-}{p^++p^-} = \tanh y .
\end{equation}

\paragraph{Relative velocity}
Rapidity is the natural relativistic velocity variable.  Suppose
we have a proton and a pion with the same rapidity at $p_{T}=0$.
Then they have no relative velocity; to see this, one just boosts
to the rest frame of one of the particles.  But these same
particles have very different energies: $E_{p}= \frac {m_{p}}{m_{\pi }} E_{\pi
}$.

\subsection{Pseudo-rapidity}

As I will now explain, the rapidity of a particle can easily be
measured in a situation where its mass is negligible, for then it
is simply related to the polar angle of the particle.

First let us define the pseudo-rapidity of a particle by
\begin{equation}
   \eta  = - \ln \tan \frac {\theta }{2} ,
\end{equation}
where $\theta $ is the angle of the 3-momentum of the particle relative
to the $+z$ axis.  It is easy to derive an expression for
rapidity in terms of pseudo-rapidity and transverse momentum:
\begin{equation}
   y = \ln \frac {\sqrt {m^{2} + p_{T}^{2} \cosh^{2}\eta } + p_{T} \sinh\eta
}{\sqrt {m^{2}+p_{T}^{2}}} .
\label{y.from.eta}
\end{equation}

In the limit that $m \ll p_{T}$, $y\to \eta $.  This accounts both for the
name `pseudo-rapidity' and for the ubiquitous use of
pseudo-rapidity in high-transverse-momentum physics.  Angles, and
hence pseudo-rapidity, are easy to measure. But it is really the
rapidity that is of physical significance: for example the
distribution of particles in a minimum bias event is
approximately uniform in rapidity over the kinematic range
available.

The distinction between rapidity and pseudo-rapidity is very
clear when one examines the kinematic limits on the two
variables.  In a collision of a given energy, there is a limit to
the energy of the particles that can be produced.  This can
easily be translated to limits on the rapidities of the produced
particles of a given mass.  But there is no limit on the
pseudo-rapidity, since a particle can be physically produced at
zero angle (or at $180^{\circ }$), where its pseudo-rapidity is infinite.
The particles for which the distinction is very significant are
those for which the transverse momentum is substantially less
than the mass. Note: from Eq.\ (\ref{y.from.eta}) it follows that
$y<\eta $ always.

\section{Diffractive Scattering}

\subsection{Kinematics}

\begin{figure}
    \centering
    \includegraphics[scale=0.8]{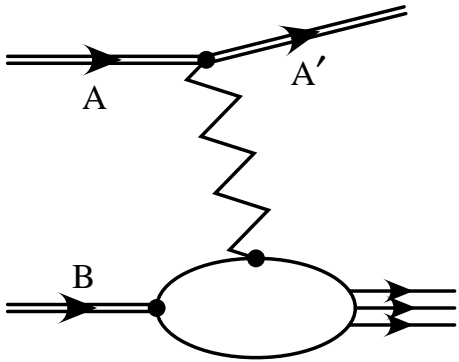}
\caption{Diffractive scattering.}
\label{fig:diffract}
\end{figure}

Let us define a single diffractive collision as one in which one
of the beam particles emerges almost unscathed from the
scattering, with only a small deflection and a small loss of
energy (Fig.\ \ref{fig:diffract}):
\begin{equation}
   p + \bar p \to  p' + X .
\label{process}
\end{equation}
For the sake of definiteness, I have assumed a proton-antiproton
collision with the proton being diffracted.  One can treat any
other collision in the same fashion.  Indeed, replacing the
incoming and outgoing protons by electrons gives us
electroproduction processes, and the low $Q^{2}$ events that are
used to measure photoproduction cross sections have exactly the
same kinematic properties as diffractive hadron collisions.

Suppose the particle $p$ is moving in the $+z$ direction, and
$\bar p$ in the opposite direction.  Then we define the
kinematics in terms of the transverse momentum $\T{p'}$ of the
diffracted hadron and a variable $x_{\Pomeron }$ defined as
$1-p'^{+}/p^{+}$.  Thus the components of $p$ and $p'$ are
\begin{eqnarray}
    p^{\mu } &=& \left( p^{+}, \frac {m^{2}}{2p^{+}}, \T 0 \right) ,
\nonumber\\
    p^{\mu } &=& \left( (1-x_{\Pomeron })p^{+},
                   \frac {m^{2}+{p'}_{T}^{2}}{2(1-x_{\Pomeron })p^{+}}, \T{p'}
           \right) .
\end{eqnarray}
It is convenient to examine the momentum transfer:
\begin{equation}
   \Delta ^{\mu } \equiv  p^{\mu }-p'^{\mu } =
           \left( x_{\Pomeron }p^{+},
                  - \frac { x_{\Pomeron }m^{2}+{p'}_{T}^{2}}{2(1-x_{\Pomeron
})p^{+}},
                  -\T{p'}
           \right) .
\end{equation}
Clearly, $x_{\Pomeron }$ represents a fractional loss of the $+$
component of momentum.  When the two protons are both moving
fast, this fraction is also the fractional energy loss:
$x_{\Pomeron } = 1-E'/E + O(1/E)$; but the definition in light cone
components converts this to a boost-invariant definition.
The invariant momentum transfer is
\begin{eqnarray}
   t = \Delta ^{2} &=& - \frac {{p'}_{T}^{2} + x_{\Pomeron
}^{2}m^{2}}{1-x_{\Pomeron }}
\nonumber\\
          &\to & -{p'}_{T}^{2} \mbox{~as $x_{\Pomeron } \to  0$}.
\end{eqnarray}
Thus at small $x_{\Pomeron }$, $t$ is an equivalent variable to the
transverse momentum of the outgoing diffracted hadron.

The diffractive region is when $x_{\Pomeron }$ is small enough that
the power law attributed to pomeron exchange dominates.  The
exact nature of the pomeron in QCD is still quite uncertain.
But from a practical point-of-view one can identify it with
whatever is exchanged between hadrons to make the approximately
constant total cross sections that are observed at high-energy.
The object exchanged between the protons in Fig.\
\ref{fig:diffract} and the system labeled $X$ is a pomeron when
$x_{\Pomeron }$ is sufficiently small.  In the Regge phenomenology of
diffractive scattering, this is associated with an $x_{\Pomeron }$
distribution of the events that is approximately
$dx_{\Pomeron }/x_{\Pomeron }$.

A more general definition of diffraction is that it consists of
events in which there are non-exponentially suppressed rapidity
gaps.  The events that are diffractive in the first definition in
fact are also diffractive according to the second definition.  As
I will show in Sect.\ \ref{sec:rap.gap}, events with small
$x_{\Pomeron }$ have a rapidity gap that grows like $\ln
(1/x_{\Pomeron })$.  The power law associated with pomeron exchange
gives a distribution of events that is approximately
$dx_{\Pomeron }/x_{\Pomeron }$, and hence approximately a constant
distribution in the size of the rapidity gap.  That is, it gives
non-exponentially suppressed rapidity gaps.

\begin{figure}
    \centering
    \includegraphics[scale=0.8]{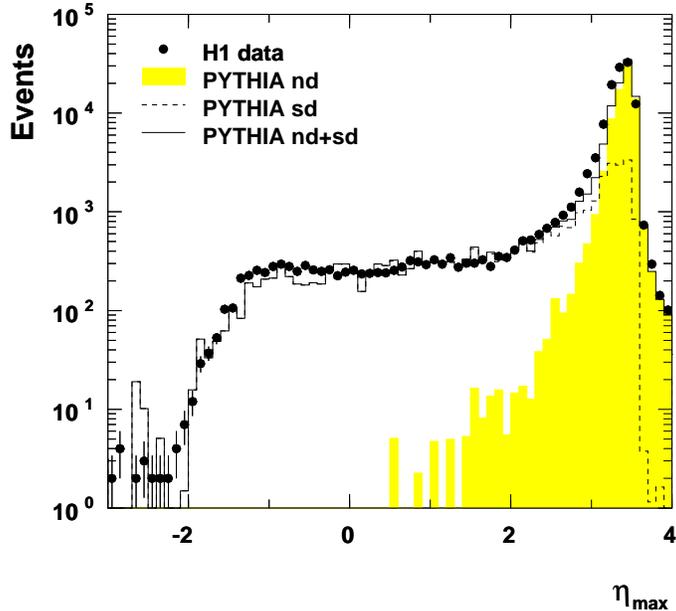}
\caption{Distribution of $\eta _{\rm max}$ in photoproduction
    at H1 \protect\cite{H1}.  There are no detected particles with
    pseudo-rapidity between $\eta _{\rm max}$ at the end of the detector
    at $\eta =3.65$
}
\label{fig:H1.rap.gap}
\end{figure}

In contrast, if one has a statistical distribution of particles,
uniform in rapidity, and if all the correlations between
particles are local in rapidity, then the probability of a
rapidity gap of size $\Delta y$ is proportional to $e^{-a\Delta y}$, for some
constant $a$. Experimental data, for example, Fig.\
\ref{fig:H1.rap.gap} indicate that for small gaps there is indeed
an approximately exponential fall-off with increasing $\Delta y$.
But for large enough gaps, more than 2 or 3 units, a presumably
different mechanism of particle production becomes significant,
and there is no longer a strong exponential drop.  Of course,
once $\Delta y$ gets close to the kinematic limit, the distribution in
$\Delta y$ must decrease again.

Strictly speaking, one should not talk about specific events as
being diffractive or non-diffractive.  Instead one should
talk about the diffractive and non-diffractive components of a
cross-section.  When one says that a specific event is
diffractive, one really means that it is in a kinematic region
(of $x_{\Pomeron }$ or rapidity gap) where diffraction is the
dominant contribution to the cross section.

\subsection{Existence of a rapidity gap}

Now the maximum value of the $+$ component of the momentum of a
hadron in the $X$ part of the final state in (\ref{process}) is
$\Delta ^{+}=x_{\Pomeron }p^{+}$.  It then follows from Eq.\ (\ref{particle})
that the maximum possible value of rapidity for any particle in
$X$ is
\begin{equation}
   y_{\rm kin.\ max} = \ln \frac {2x_{\Pomeron }E}{m} ,
\label{kin.max}
\end{equation}
where $E$ is the beam energy (assumed to be much larger than the
proton mass).

For a numerical example, consider a diffractive process with
an incoming proton of energy 900 GeV (as at the Fermilab
Tevatron), and with $x_{\Pomeron }=0.01$, a value typically
considered to be safely in the diffractive region. The incoming
proton itself has rapidity 7.6. From Eq.\ (\ref{kin.max}) the
maximum possible rapidity of pions in the event is 4.9. There is
therefore a rapidity gap on the proton side of the event.  Since
there are normally many particles in the event, which share the
momentum, the edge of the rapidity gap, $y_{\rm max}$, will
typically be a unit or more lower.

\begin{figure}
    \centering
    \includegraphics[scale=0.7]{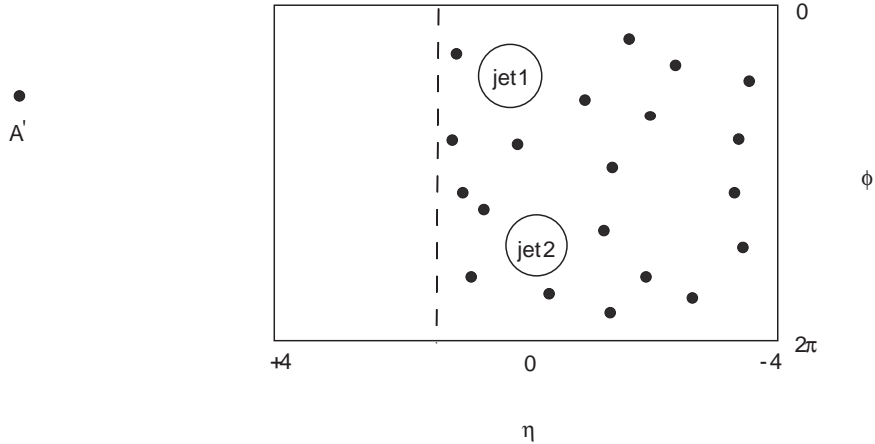}
    \\*[5mm]
\caption{Plot of the distribution of final-state particles in a
         typical diffractive event is uniform in $\phi $ and $y$
         except for an isolated forward proton.}
\label{fig:event}
\end{figure}

The distribution of the soft particles in a hadronic collision is
approximately uniform in rapidity.  So when the final-state
particles are plotted in azimuth $\phi $ and rapidity $y$, a
diffractive event therefore looks like Fig.\ \ref{fig:event}.
When $x_{\Pomeron }$ is decreased, the value of $y_{\rm kin.\ max}$
also decreases, and hence the size of the rapidity gap increases.

\subsection{Computation of $x_{\Pomeron }$ from the non-diffracted
    part of the final state}

In principle, $x_{\Pomeron }$ can be either measured from the final
state hadron: $x_{\Pomeron } = 1-p'^{+}/p^{+}$ or from the rest of the
final state:
\begin{equation}
   x_{\Pomeron } = \sum _{i\epsilon X} \frac {k_{i}^{+}}{p^{+}} ,
\label{xP.from.X}
\end{equation}
where the sum is over all particles in the remaining part $X$ of
the final state.

To measure $x_{\Pomeron }$ from the diffracted hadron requires first that
there be elements of the detector (e.g., Roman pots) close to the
beam-pipe, which is not always the case.  Moreover, if $x_{\Pomeron }$ is
very small, the hadron's energy needs to be known very precisely
to obtain a useful accuracy on $x_{\Pomeron }$; if the measurement is not
sufficiently accurate, one cannot distinguish the value from
zero.  (Values of $x_{\Pomeron }$ in the range $10^{-4}$ to $10^{-2}$ are often
discussed.)  So if a useful measurement of $x_{\Pomeron }$ is not
available, one must measure the rest of the
hadronic final-state, and deduce $x_{\Pomeron }$ by using Eq.\
(\ref{xP.from.X}).

In principle, the use of Eq.\ (\ref{xP.from.X})
requires knowledge of all the (typically many) particles in $X$.
One might imagine that poor information on $X$, for example from a
lack of calorimetry in the backward part of the detector, would
prevent such a measurement from being practical.

In fact, as I will now show, the value of the right-hand-side of
Eq.\ (\ref{xP.from.X}) is dominated by the particles of the
largest $k_{T}$ (i.e., jet particles), and by the particles that are
closest in rapidity to the diffracted hadron, i.e., closest to
the edge of the rapidity gap.

We simply write each $k_{i}^{+}$ in Eq.\ (\ref{xP.from.X}) in terms of
rapidity and transverse momentum to obtain
\begin{equation}
   x_{\Pomeron } = \sum _{i\epsilon X} \frac {E_{iT} \e^{y_{i}}}{2E} .
\label{xP.from.X1}
\end{equation}
Here I have approximated $p^{+}$ by $E\sqrt 2$, with $E$ being the
energy of the incoming proton, which is assumed to be large.  The
transverse momentum appears in form of the transverse energy
$E_{iT} = \sqrt {\strut k_{iT}^{2}+m^{2}}$.
We next recall that particles in a hadronic scattering event are
typically uniformly distributed in rapidity between some
kinematic limits, and have a steeply falling spectrum in
transverse momentum.  The exponential of rapidity in Eq.\
(\ref{xP.from.X1}) therefore implies that the sum is dominated by
those particles of the highest rapidity.

The exception to this statement is given by particles that result
from a hard scattering and that therefore have large transverse
momentum. Hence the general statement is that the sum in Eq.\
(\ref{xP.from.X1}) is dominated by (a) particles with the largest
rapidity (i.e., closest to the edge of the rapidity gap), and (b)
particles with high transverse momentum. This means that the
measurement can be effectively be made from
\begin{equation}
   x_{\Pomeron } \simeq
   \sum _{{\rm particles\ with\ highest\ }y \atop {\rm particles\ with\ large\
}E_{T}}
   \frac {E_{iT} \e^{y_{i}}}{2E} .
\label{approx.xpom}
\end{equation}
Hence a lack of detector coverage on the opposite side to the
diffracted proton does not greatly affect the accuracy of the
measurement.

Exactly this method is used in a corresponding problem in
deep-inelastic scattering, where it is known as the
Jacquet-Blondel \cite{JacquetBlondel} method.  There it is
desired to measure the scaling variable $y$ (which has no
relation to rapidity).  This variable is defined in terms of the
scattered lepton in exactly the way in which our $x_{\Pomeron }$ is
defined in terms of the diffracted proton.  By momentum
conservation, $y$ can be computed from the hadronic final state.
The Jacquet-Blondel method is in regular use in deep-inelastic
experiments at the HERA collider, and it provides useful
accuracy.  It is therefore evident that the same method applied
to diffraction may well provide a practical method for measuring
$x_{\Pomeron }$.\footnote{
    One counterargument is that when one measures $y$
    in deep-inelastic scattering, the
    particles dominating the sum corresponding to the one
    in Eq.\ (\ref{approx.xpom}) are
    those with large transverse momentum, whereas in diffractive
    hadron scattering, the soft particles, those in the
    underlying event, are important. In diffractive hadron
    scattering one will measure all the relevant low transverse
    momentum particles only if the tracking is very good.
}

\subsection{Scaling variable $\betaH $ relative to pomeron}

\begin{figure}
    \centering
    \includegraphics[scale=0.7]{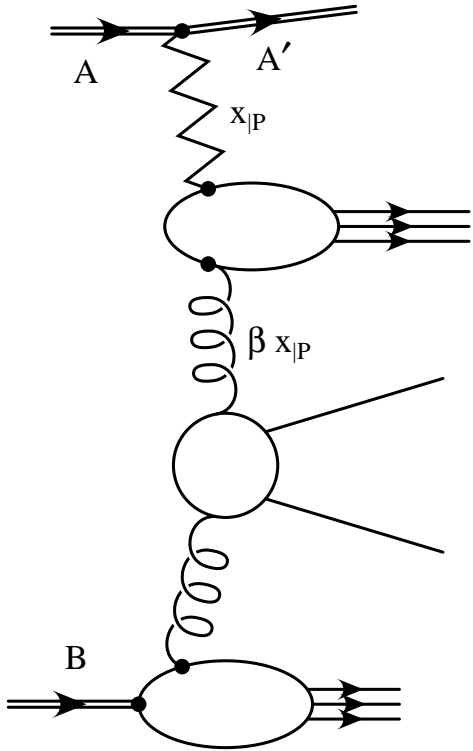}
    \\[6mm]
\caption{Diffractive hard scattering.  A parton inside the
         pomeron undergoes a scattering scattering with a parton
         out of the antiproton.  }
\label{fig:diff.hard}
\end{figure}

When one has not only diffraction but also a hard scattering, there is
another important kinematic variable that can be measured by a similar
method.  A closely related variable is the one called $\beta$ in
diffractive deep-inelastic scattering.  I have added a subscript `H'
to distinguish my definition, with `H' denoted `hard scattering'.
The definition and its physical
interpretation can be explained by considering a diffractive hard
collision treated by the QCD factorization formula, as formulated
by Ingelman and Schlein \cite{IS}. The situation is represented
in Fig.\ \ref{fig:diff.hard}.  There the hard part (i.e., the
large $p_{T}$ subprocess) of the collision is treated as being due
to a collision of a parton from the antiproton with a parton out
of the exchanged pomeron.  The parton in the pomeron has a
distribution that is a function of its momentum fraction, $\betaH$, in
the pomeron.  Now fractions here are defined to be fractions of
the $+$ component of momenta.  So we can measure $\betaH $ from the
hadronic final state by
\begin{equation}
   \betaH  = \frac {x_{\rm jet}}{x_{\Pomeron }} ,
\end{equation}
where $x_{\rm jet}$ is defined in the same way as $x_{\Pomeron }$,
except that the sum is restricted to the particles that are
generated by the hard scattering subprocess, i.e., the large
transverse momentum particles.  Thus
\begin{equation}
   \betaH  =
    \frac {\displaystyle \sum _{{\rm particles\ with\ large\ }E_{T}} \frac {
E_{iT} \e^{y_{i}}}{2E} }{\displaystyle \sum _{{\rm all\ particles}} \frac {
E_{iT} \e^{y_{i}}}{2E} } .
\end{equation}
The sum in the denominator is effectively over just the particles
with the largest rapidity and/or transverse energy.
The importance of measuring cross sections differential in $\betaH $ is
that it gives a direct probe of the parton densities in the
pomeron.

I use the name `jet' in the symbol $x_{\rm jet}$, in view of the
fact that an important application is to events containing high
$p_{T}$ jets.  In that case the sums are actually over hadrons.
However, it is important to observe that the same principles
apply to any other hard process, like Drell-Yan.  Then the sums
are over all particles of the specified kinematics, hadronic or
leptonic.

{\rm Important note about notation: } In diffractive deep-inelastic
scattering, it is common to define a variable $\beta \equiv x_{\rm
  bj}/x_{\Pomeron}$.  In an event that is generated by the lowest-order
mechanism of elastic electron-quark scattering, it can readily be
verified that $\beta=\betaH$.  This is in fact a motivation for the
definition of $\beta$.

Moreover, although the motivation for the definition of $\betaH $, and
many of its uses, are associated with the factorization formula
(the `Ingelman-Schlein model'), the reader should observe that
the definition of $\betaH $ is completely independent of the model.
The expression for $\betaH $ involves only properties of measurable
hadrons (and leptons) in the collision.  It does not involve
the unobserved partons that are the mechanism by which the
process is supposed to be generated.  This observation is
particularly important in the light of the fact that the
factorization formula for diffractive hard scattering has not
been proved from QCD, unlike the related formula for inclusive
hard scattering.

\subsection{Relation between rapidity gaps and $x_{\Pomeron }$}
\label{sec:rap.gap}

Much diffractive data is presented as a function of the
position of the edge of the pseudo-rapidity gap rather than as a
function of $x_{\Pomeron }$.  So it is useful to obtain an
approximate relation between the two variables, in order to be
able to compare different sets of data.  I will present results
both for the gaps in rapidity and in pseudo-rapidity.

\subsubsection{Estimation of the size of the gap}

\begin{figure}
    \centering
    \includegraphics[scale=0.7]{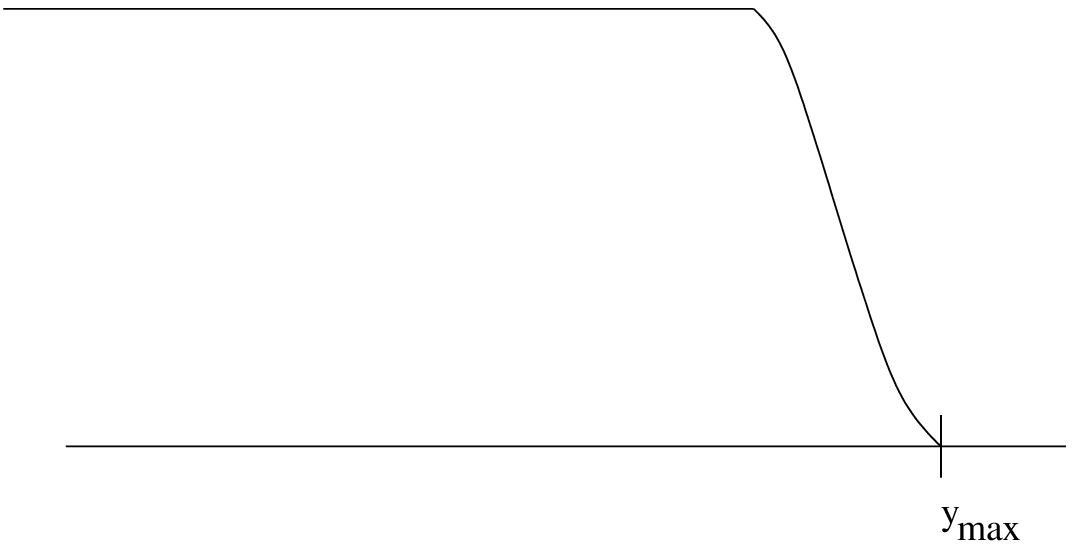} \\
           (a)  \\[5mm]
    \includegraphics[scale=0.7]{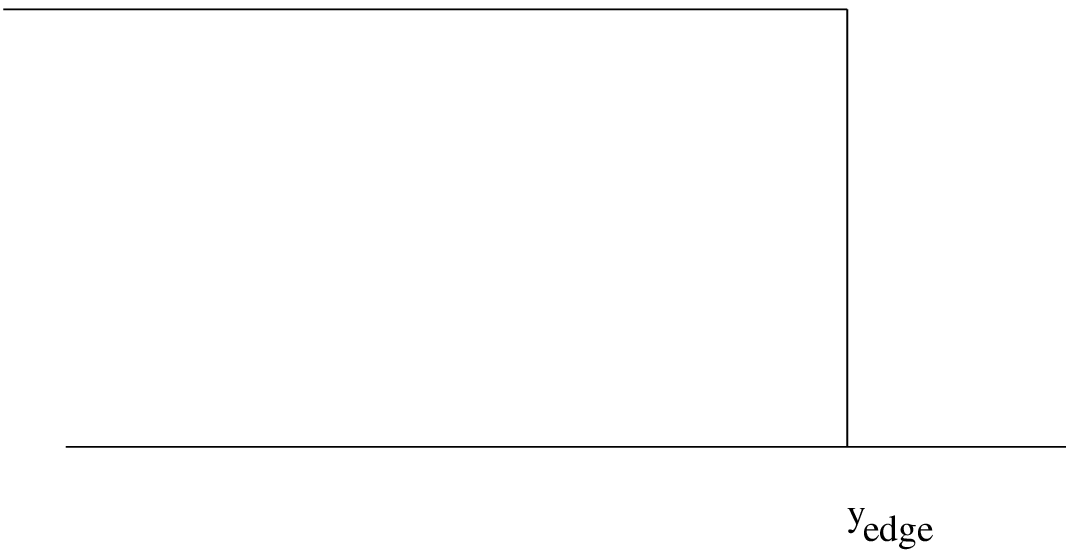} \\
           (b)  \\[5mm]
\caption{(a) True distribution of particles in a diffractive event.
         (b) Theta-function approximation to this distribution.}
\label{fig:dN.dy}
\end{figure}

In a typical event, the distribution of particles as a function
of $y$ is supposed to look like Fig.\ \ref{fig:dN.dy}(a).  To
estimate the relation between the edge of the rapidity gap and
$x_{\Pomeron }$, I approximate the true distribution by a theta
function distribution, Fig.\ \ref{fig:dN.dy}(b).  In the
approximate distribution we have a uniform distribution, for
values of $y$ less than a value I will call $y_{\rm edge}$, and no
particles above $y_{\rm edge}$. The height of the distribution I
will call $dN/dy$.  From the distribution of hadrons in real
diffractive events \cite{SDE}\footnote{
   There is also earlier data from the ISR \cite{ISR}.
}
I estimate that an appropriate
value for $y_{\rm edge}$ is about one unit less than the actual end
of the rapidity gap, in order to obtain the same value for
$x_{\Pomeron }$.

Next, I split the computation of $x_{\Pomeron }$ into a sum over the
high $p_{T}$ particles (e.g., associated with jets), and a sum over
the low $p_{T}$ particles (which comprise what is often termed the
`underlying event').  For the particles in the underlying event, I
replace their transverse energy by its mean value $\langle E_{T}\rangle$.
This gives
\begin{eqnarray}
    x_{\Pomeron } &\simeq& \betaH x_{\Pomeron }
        + \int _{-\infty }^{y_{\rm edge}} dy \, \e^{y} \, \frac {dN}{dy} \,
\frac {\langle E_{T}\rangle }{2E}
\nonumber\\
    &=& \betaH x_{\Pomeron }
        + \e^{y_{\rm edge}} \, \frac {dN}{dy} \, \frac {\langle E_{T}\rangle
}{2E} .
\end{eqnarray}
Here, I have extended the integral over $y$ to $-\infty $, since, in
the underlying event, only the hadrons closest to the edge of the
rapidity gap are important in the calculation. Hence
\begin{eqnarray}
    \ln \frac {1}{x_{\Pomeron }}
  &\simeq&
    - \ln \frac {1}{1-\betaH }
    - y_{\rm edge} - \ln \frac {dN}{dy} + \ln \frac {2E}{\langle E_{T}\rangle }
\nonumber\\
  &=&
    -y_{\rm max} - \ln \left( \frac {1}{1-\betaH } \right)
    + \left[
         \ln \frac {2E}{\langle E_{T}\rangle }
       + y_{\rm max} - y_{\rm edge}
       -\ln \frac {dN}{dy}
    \right] ,
\label{xpom.from.ymax}
\end{eqnarray}
with the term in square brackets being approximately independent
of the event's kinematic parameters, $y_{\rm max}$ and $\betaH $.

\subsubsection{Numerical values}

For a relation between $x_{\Pomeron }$ and the maximum
measured {\em pseudo}rapidity, let us rewrite Eq.\
(\ref{xpom.from.ymax}) as
\begin{equation}
   \ln \frac {1}{x_{\Pomeron }} =  C - \eta _{\rm max}
   - \ln \frac {1}{1-\betaH } + \ln \frac {E}{900 \, {\rm GeV}},
\label{xpom.from.eta.max}
\end{equation}
with
\begin{equation}
   C = \eta _{\rm max} - y_{\rm edge}
       + \ln \frac {1800 \, {\rm GeV}}{\langle E_{T}\rangle } - \ln \frac {
dN}{dy} .
\end{equation}
Now $C$ is a pure number that should be approximately independent
of any of the parameters of the collision. So we can try to
estimate it from the CDF data on single diffraction \cite{SDE}.
Their plots for the number distribution of {\em charged} tracks
as a function of pseudo-rapidity are averaged over all events.
Since different events have different values of $x_{\Pomeron }$, the
plot need not exhibit a rapidity plateau even if individual
events do so. Moreover the rise of $dN/dy$ from zero to its
plateau will be spread out.

{}From the data, I estimate that at $\sqrt s=1800\,{\rm GeV}$
\begin{eqnarray}
         \frac {dN}{d\eta }   &\approx & 2,               \nonumber\\
  \eta _{\rm max} - y _{\rm  edge} &\approx & 1.
\label{SDE}
\end{eqnarray}
In addition I use the value \cite{CDF.min.bias} $\langle E_{T}\rangle
\approx  0.3\, {\rm GeV}$.
These values give $C=8.5$.

Eq.\ (\ref{xpom.from.eta.max}) is, of course, not exact on an
event-by-event basis. However, it should be useful on average, with
the real data being smeared around the values related by the formula,
and of course the numerical value `8.5' should be adjusted by a fit to
ordinary single diffraction data.

There is indeed such data.  CDF in Ref.\
\onlinecite{CDF-hard-diff} give a plot of $\eta _{\rm max}$ against
$x_{\Pomeron }$  for their diffractive dijet data. Unfortunately, it
disagrees quite badly with the above formula, both as to
normalization and slope. This needs to be investigated.  For
example, the numbers I quoted above in Eq.\ (\ref{SDE}) may be
wrong, and my formulae have made no allowance for the fact that
the measured $\eta _{\rm max}$ applies to particles above some value of
$E_{T} \approx  400 \, {\rm MeV}$.  The true $\eta _{\rm max}$ may be rather
larger.

\section*{Acknowledgments}

This work was supported in part by the U.S.\ Department of Energy
under grant number DE-FG02-90ER-40577.
I would like to thank Mike Albrow, Andrew Brandt, and Kristal
Mauritz for discussions.


\end{document}